# TRIPLE-GEM PERFORMANCE IN He-BASED MIXTURES

A. BONDAR, A. BUZULUTSKOV, L. SHEKHTMAN, A. VASILJEV

*Budker Institute of Nuclear Physics, Novosibirsk 630090, Russia*

The performance of triple-GEM detectors in He+$N_2$ gas mixtures is studied in the range of 1-10 atm. The results obtained are relevant in the field of minimization of ionic space-charge effect in the TPC and neutron detection.

The unique advantage of He-based mixtures (and those of Ne) is that the ion mobilities in He (and Ne) are by an order of magnitude higher than those in Ar (see Table 1)[1]. The unique advantage of Gas Electron Multipliers (GEMs)[2] is that they can operate at high gains in pure noble gases[3], and in particular in pure He and Ne[4]. Accordingly, using He- and Ne-based mixtures in multi-GEM structures could help to solve the problem of ion accumulation in the GEM-based TPCs. In addition, He-based detectors are relevant in the field of neutron detection[4]. However, the electron drift velocity in pure He and Ne is rather low, which may decrease the rate capability of the device. The latter problem might be solved using mixtures with $N_2$ additive: in particular very fast signals were observed in Ar+$N_2$ in multi-GEM structures[3]. In this paper, we study the performance of triple-GEM detectors in mixtures of He with $N_2$ at high pressures.

| gas | He | | Ne | | Ar | |
|---|---|---|---|---|---|---|
| ion | $He^+$ | $He_2^+$ | $Ne^+$ | $Ne_2^+$ | $Ar^+$ | $Ar_2^+$ |
| $K_0, \frac{cm^2}{V \times s}$ | 10.4 | 16.7 | 4.1 | 6.5 | 1.5 | 1.86 |

Table 1: Ion mobilities in He, Ne and Ar[1].

The experimental setup and procedures are described elsewhere[4]. Fig. 1 (left) shows gain-voltage characteristics of the triple-GEM in He+1%$N_2$ and He+10%$N_2$ in the range of 1-10 atm. Rather high gains are obtained in these mixtures, exceeding $10^6$ and $10^4$, respectively, presumably due to the fact that these mixtures are actually Penning mixtures having an enhanced ionization efficiency.

One can see from Fig. 1 (right, top) that the rise-time of the pulse is about 100 ns, corresponding to the width of the anode pulse of 50 ns. This should be compared to the pulse width in pure He, of 150 ns [5]. Fig. 1 (right, bottom) illustrates the ion feedback effect in Ar+10%N$_2$: at a gain of about 5000 the primary pulse is accompanied by a number of after-pulses. Assuming that ion feedback takes place between the last and last-but-one GEM elements, one may estimate the ion mobility in this mixture: it amounts to 15 cm$^2$/Vs which is by an order of magnitude higher than that in Ar. Comparing the data with Table 1, one may conclude that the principle ions in an avalanche might be molecular ions He$_2^+$.

In conclusion, it is possible to considerably reduce the effect of ionic space charge in the TPC volume using He-based mixtures due to higher ion mobility. Adding a small amount of nitrogen to He significantly decreases the width of the anode pulse. Ion mobility in a given mixture in multi-GEM structures can be estimated using the pulse-shape analysis.

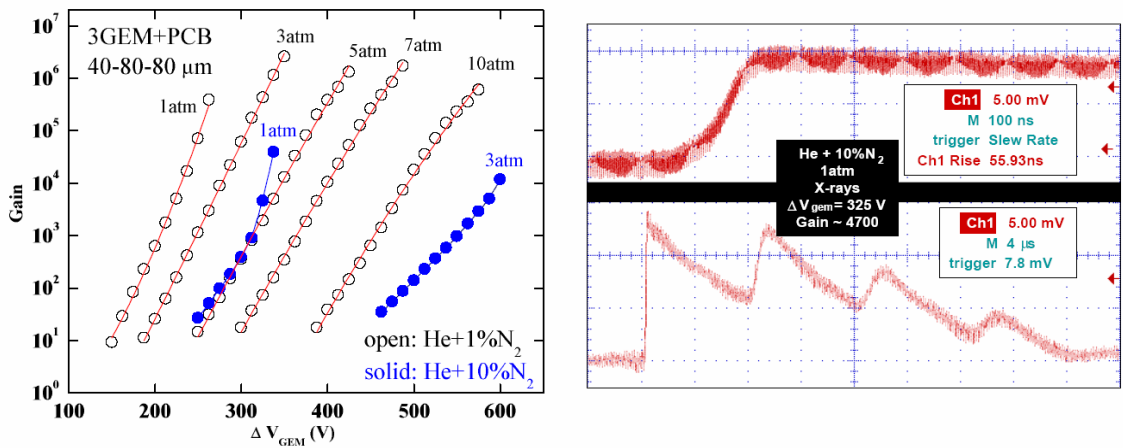

Figure 1: Gain-voltage characteristics (left) and anode signals using charge-sensitive amplifier (right), from the triple-GEM in He+N$_2$.

## References

1. A. A. Radtsig, B. M. Smirnov, in: B.M. Smirnov (Ed.), Plasma Chemistry, Energoizdat, Moscow, v. 11, 1984, p. 170.
2. F. Sauli, Nucl. Instr. and Meth. A 386 (1997) 531.
3. A. Buzulutskov et al., Nucl. Instr. and Meth. A 443 (2000) 164.
4. A. Bondar et al., Nucl. Instr. and Meth. A 493 (2002) 8.
5. V. Aulchenko et al., Nucl. Instr. and Meth. A 513 (2003) 256.